\documentclass[final,5p,times,sort&compress]{elsarticle}

\usepackage{graphicx}
\usepackage{amssymb}
\usepackage{lineno}
\usepackage{color}
\usepackage{url}
\usepackage{bm}

\begin{document}

\begin{frontmatter}

\title{On the atomistic energetics of carbon nanotube collapse from AIREBO potential}

\author[Ume_add]{Yoshitaka Umeno}
\author[YY_add]{Yu Yachi}
\author[MS_add]{Motohiro Sato}
\author[HS_add]{\corref{cor1}Hiroyuki Shima}

\address[Ume_add]{Institute of Industrial Science, The University of Tokyo, 4-6-1 Komaba, Meguro-ku,Tokyo 153-8505, Japan}
\address[YY_add]{Division of Socio-Environmental Engineering, Graduate School of Engineering, Hokkaido University, 
N13-W8, Kita-ku, Sapporo, Hokkaido 060-8628, Japan}
\address[MS_add]{Division of Socio-Environmental Engineering, Faculty of Engineering, Hokkaido University, 
N13-W8, Kita-ku, Sapporo, Hokkaido 060-8628, Japan}
\address[HS_add]{Department of Environmental Sciences, University of Yamanashi, 4-4-37, Takeda, Kofu, Yamanashi 400-8510, Japan}

\cortext[cor1]{To whom correspondence should be addressed: Hiroyuki Shima (Email: hshima@yamanashi.ac.jp)}

\begin{abstract}
Molecular dynamics simulations based on the adaptive intermolecular reactive empirical bond order (AIREBO) were performed to probe hydrostatic pressure induced collapse of single-walled and double-walled carbon nanotubes. It was unveiled that the torsion term, which is a specific potential component involved in the AIREBO scheme, plays a vital role in stabilizing fully collapsed cross-sections of the carbon nanotubes. Evolution of the cross-sectional deformation along the loading-unloading curve was also elucidated, showing strong dependence on the presence of a structural defect on the outer carbon wall.
\end{abstract}



\end{frontmatter}

\section{Introduction}

Radial buckling of carbon nanotubes is one of the most important deformation modes
that give rise to dramatic changes in their physical properties \cite{ShimaMater2012}.
It is mainly a consequence of the hollow cylindrical structure of carbon nanotubes;
the radial stiffness of the monoatomic cylinder, made by rolling up a graphene sheet,
is much less than its remarkably high axial stiffness.
Such the anisotropy in the stiffness allows cross-sectional buckling of carbon nanotubes
when they are subjected to external high pressure
or suffered from mechanical instability \cite{ShimaBook2012}.

A primary example of the radial buckling mode
is the one observed in pressurized many-walled carbon nanotubes.
This mode is characterized by wavy-shaped cross-sections \cite{ShimaNTN2008,ShimaPRB2010,UmenoPE2014},
which result from mechanical instability of outer walls that are reinforced by inner walls.
Another example of radial buckling mode occurs in single-walled carbon nanotubes (SWNTs) with large radius.
In the latter systems, 
loss of reinforcing inner walls makes the circular cross-section 
be energetically unstable.
Beyond a threshold radius, therefore,
they undergo spontaneous collapse in the radial direction
\cite{ChopraNature1995,BenedictCPL1998,GHGaoNTN1998,PantanoJMPS2004,ElliottPRL2004,
HJLiuAPL2004,BLiuPRB2004,DYSunPRB2004,XHZhangPRB2004,ZWWangJPCB2004,TTangJAP2005,
SLZhangPRB2006,JXiaoNTN2007,MottaAdvMater2007,TChangNanoLett2010,
CHZhangACSNano2012,YLiPCCP2014,HMaACSNano2014,CerqueiraCarbon2014}.
A fully collapsed cross-section is composed of two parallel line segments 
whose ends are capped by two highly strained circular edges.
Similar collapsing behavior was found to occur in double-wall carbon nanotubes (DWNTs)
\cite{XYangAPL2006,AnisPRB2012,AguiarJPCC2012,AlencarCarbon2017}.

From the physics perspective,
the significance of nanotube collapse lies in that it may cause 
semiconductor-metal transition \cite{BarbozaPRL2008,GiuscaNanoLett2008}, 
optical response change \cite{ThirunavukkuarasuPRB2010}, 
magnetic moment quenching \cite{DinizPRB2010},
as well as interwall sp$^3$ bonding
between adjacent walls \cite{FonsecaPRB2010,SakuraiPhysicaE2011,
ZHXiaPRL2007,ByrnePRL2009,FilleterAdvMater2011,YYZhangJAP2011}.
Furthermore, nanotube collapse is expected to provide a way of synthesizing 
closed-edged bilayer graphene nanoribbons \cite{CZhangACSNano2012,DHChoiSciRep2013},
which hold promise as building blocks in high-performance nanoelectronics
because of the fascinating mechanical
\cite{MottaAdvMater2007,XHZhongNTN2012,VilatelaCarbon2011}
and electrical properties \cite{SengaAPL2012,DHChoiSciRep2013}.
Against the backdrop, deep insight into an atomistic-scale mechanism of nanotube collapse
will provide important clues for the discovery of novel physical properties of carbon nanotubes
and their potential advantages in the field of nanoengineering.

In the present work, we perform molecular dynamics (MD) simulations
of collapsing SWNTs and DWNTs with and without structural defects.
Our MD simulation code is based on a broadly used interatomic potential
(called AIREBO potential \cite{StuartJCP2000}; see \S \ref{subsec_AIREBO}),
which makes it possible to express gradual changes from covalent to non-covalent bonds
and vice versa in accord with interatomic distance as well as
local atomic configuration.
The main purpose of this work is to clarify what parts of AIREBO potential components 
should be dominant for driving and finally stabilizing the collapsed cross-section
of nanotubes under high hydrostatic pressure.
We also establish the pathway of cross-sectional deformation along the loading-unloading curve,
paying attention to the way how the presence of structural defects in the wall
gives feasible changes in the deformation pathway.

\section{Method}

\subsection{Atomistic model of CNTs}

\begin{figure}[ttt]
\centering
\includegraphics[width=7.5cm]{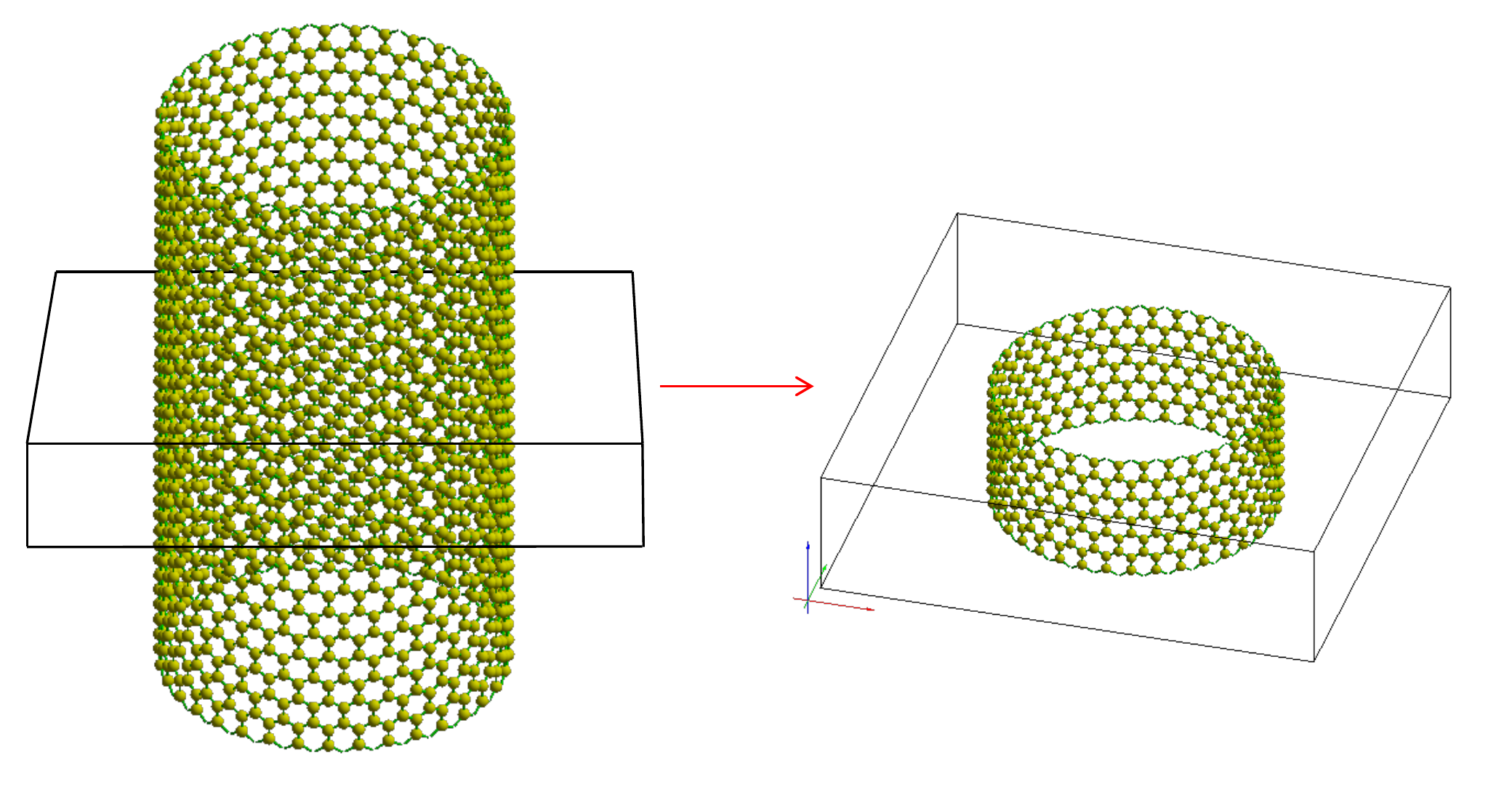}
\caption{(Left) Overall view of a zigzag SWNT with chiral index of (36,0).
(Right) A unit cell surrounded by the simulation box having 12.78 \AA$\;$ in height.}
\label{fig_unitcell}
\end{figure}

In this study we deal with zigzag SWNTs with the chiral index of (36,0)
and DWNTs with (27,0)@(36,0). 
Figure \ref{fig_unitcell} presents the simulation cell of a (36,0) SWNT.
The unit cell is arranged such that the tube axis is parallel to the $z$ axis;
the height of the simulation box depicted in Fig.~\ref{fig_unitcell} by solid lines equals three times
the unit length of the (36,0) SWNT
along the tube axis (12.78 \AA), in which 432 carbon atoms are contained.
The periodic boundary condition is imposed only in the $z$ direction.

We also deal with SWNTs and DWNTs endowed with a structural defect,
called the Stone-Wales (SW) defect \cite{StoneCPL1986}.
A SW defect is formed by a $\pi/2$ rotation of a C-C bond, which transforms four hexagonal carbon rings 
into two pentagons and two heptagons.
This defect structure is a metastable state,
in the sense that the system needs to overcome an activation barrier of several eV to realize it.
It was suggested that the presence of SW defects gives an impact on the mechanical stability
of collapsing carbon nanotubes \cite{CCLing2012}.
In our actual simulations, one SW defect is imposed to each nanotube sample
at the midpoint in the height of the unit cell.

\subsection{Mechanical energy}\label{subsec_AIREBO}

\begin{figure}[ttt]
\centering
\includegraphics[width=9.5cm]{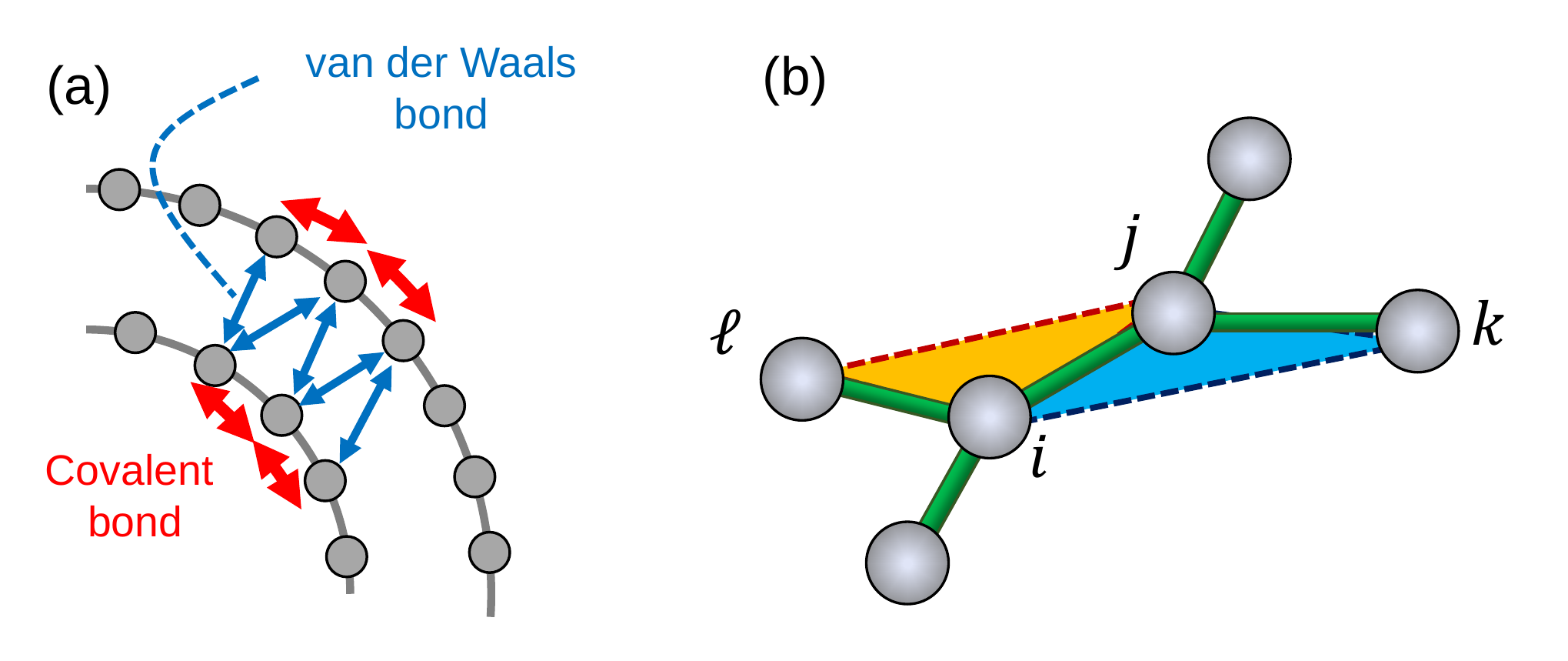}
\caption{(a) Two different types of carbon-carbon bonds in a DWNT.
Neighboring carbon atoms are covalently bonded
if they are contained in the same graphitic wall;
otherwise, they are bonded by van der Waals forces
defined by the Lennard-Jones potential.
(b) Local atomic configuration adjacent to the covalent bond 
connecting the atoms $i$ and $j$ in a graphene sheet.
The dihedral angle between the two triangles,
marked by $i-j-k$ and $i-j-\ell$,
determines the torsion energy of the C-C-C-C chain that bridges 
the atom $k$ with its third nearest neighbor $\ell$.}
\label{fig_bonding}
\end{figure}

Forces exerted on atoms are calculated by 
implementing the adaptive intermolecular reactive empirical bond order (AIREBO) 
interatomic potential \cite{StuartJCP2000}.
The mechanical energy of a given nanotube is represented by
\begin{equation}
E = \frac12 \sum_i \sum_{j\ne i} 
\left( 
E_{ij}^{\rm REBO} + E_{ij}^{\rm LJ} + \sum_{k\ne i,j} \sum_{\ell \ne i,j,k} E_{kij\ell}^{\rm tors}
\right),
\label{eq_005}
\end{equation}
with $i,j,k$ and $\ell$ being atom numbering.
The three terms involved in the parenthesis in Eq.~(\ref{eq_005})
represent different energy components as explained below in turn.

The first term, written by $E_{ij}^{\rm REBO}$,
is the second generation reactive empirical bond order (2g-REBO) potential
\cite{BrennerJPCM2002}.
It represents the potential energy of covalent bonds between nearest-neighbor carbon atoms,
which are highlighted by thick arrows (colored in red) in Fig.~\ref{fig_bonding}(a).
The value of $E_{ij}^{\rm REBO}$ depends not only on the interatomic distance between $i$th and $j$th atoms
but also local atomic configuration around the two atoms.
In general, the 2g-REBO potential 
makes it possible to evaluate with high accuracy
the in-plane elasticity and bending rigidity of 
a graphene layer that consists of sp$^2$ bonds.

The second term in the parenthesis, $E_{ij}^{\rm LJ}$, accounts for 
the van der Waals (vdW) interaction between atoms that belong to different graphitic layers
facing each other, as marked by thin arrows (colored in blue) in Fig.~\ref{fig_bonding}(a).
The dependence of $E_{ij}^{\rm LJ}$ on the interatomic distance
was set to be the same as that of the ordinary Lennard-Jones (LJ) 6-12 interatomic potential,
as implied by the superscript;
the well depth and equilibrium distance of the LJ potential were set to be
2.84 meV and 3.4 \AA, respectively.

The third term, $E_{kij\ell}^{\rm tors}$, the so-called ``torsion term",
is derived from the energy required for 
torsional displacement of a carbon atom relative to its third nearest neighbor 
in the same graphitic layer.
Figure \ref{fig_bonding}(b) illustrates the way of evaluating
the torsion term from local atomic configuration in a graphene layer.
Note that a covalent bond that connects the atoms $i$ and $j$
is associated with four dihedral angles,
one of which is the angle between the two triangular planes depicted in Fig.~\ref{fig_bonding}(b).
Each of the four dihedral angles accounts for an interaction 
between one atom (e.g., labeled by $k$) and one of its third nearest neighbors ($\ell$).
A dihedral angle $\omega_{kji\ell}$ is either 0 or $\pi$ if all the four relevant atoms lie in a flat plane;
specifically in the flat configuration, $E_{kij\ell}^{\rm tors}$ takes a maximum (or minimum) value
at $\omega_{kji\ell}=0$ (or $\pi$)
that corresponds to {\it cis-} ({\it trans-})geometry of the C-C-C-C chain.

Among the three constituent terms in AIREBO,
the working of the first two, 
$E_{ij}^{\rm REBO}$ and $E_{ij}^{\rm LJ}$,
on mechanical properties of collapsing carbon nanotubes
has been intensively investigated.
In particular, $E_{ij}^{\rm LJ}$ is known to play a crucial role in 
radial collapse of nanotubes,
because the attraction and repulsion between opposing graphene walls 
are believed to stabilize collapsed cross-sections.
The remaining one, $E_{kij\ell}^{\rm tors}$, however, has not been paid
sufficient attention thus far.
This fact motivated the authors to examine 
the impact of torsional atomic interaction represented by this term
on the nanotube collapse.

\subsection{Torsion term in AIREBO}

This subsection describes certain details 
in the mathematical expression of $E_{kij\ell}^{\rm tors}$
for later use.
Suppose the local configuration of nearby six atoms, as depicted in Fig.~\ref{fig_bonding}(b),
for which the relative position of the $i$th atom locating at $\bm{r}_i$
with respect to the $j$th atom at $\bm{r}_j$ can be represented by $\bm{r}_{ij} \equiv \bm{r}_i - \bm{r}_j$.
Then $\omega_{kji\ell}$ is defined by the angle between the plane spanned by the two vectors
$\bm{r}_{ik}$ and $\bm{r}_{ij}$
and the plane spanned by 
$\bm{r}_{ij}$ and $\bm{r}_{j\ell}$,
satisfying the relation: 
\begin{equation}
\cos \left( \omega_{kji\ell} \right) = 
\frac{\bm{r}_{ik} \times \bm{r}_{ij}}{\left| \bm{r}_{ik} \times \bm{r}_{ij} \right|}
\cdot
\frac{\bm{r}_{ij} \times \bm{r}_{j\ell}}{\left| \bm{r}_{ij} \times \bm{r}_{j\ell} \right|}.
\end{equation}
Following Ref.~\cite{StuartJCP2000}, $E_{kij\ell}^{\rm tors} $ is expressed by
\begin{equation}
E_{kij\ell}^{\rm tors} =
\sum_i
\sum_{j\ne i}
\sum_{k\ne i,j}
\sum_{\ell\ne i,j,k}
W_{kji\ell}
V^{\rm tors} \left( \omega_{kji\ell} \right),
\label{eq_123}
\end{equation}
where
\begin{equation}
W_{kji\ell} =
w_{ki} \left( r_{ki} \right)
w_{ij} \left( r_{ij} \right)
w_{j\ell} \left( r_{j\ell} \right),
\label{eq_123a}
\end{equation}
and
\begin{equation}
V^{\rm tors} \left( \omega_{kji\ell} \right)
=
 \varepsilon_{kji\ell}
\left[ 
\frac{256}{405} \cos^{10} \left( \frac{\omega_{kji\ell}}{2} \right)
-\frac{1}{10}
\right].
\label{eq_125}
\end{equation}
In Eq.~(\ref{eq_123a}), $w_{ij}(r_{ij})$ is a bond weight function
that permits the identification of atom pairs as bonded ($w_{ij}=1$),
nonbonded ($w_{ij}=0$), or partially dissociated ($0<w_{ij}<1$),
depending on the interatomic distance $r_{ij}$ \cite{StuartJCP2000}.
In our MD simulations, the barrier height $\varepsilon_{kji\ell}$ in Eq.~(\ref{eq_125})
was set to be 0.3079 eV in accord with Ref.~\cite{StuartJCP2000}.

\begin{figure}[ttt]
\centering
\includegraphics[width=8.7cm]{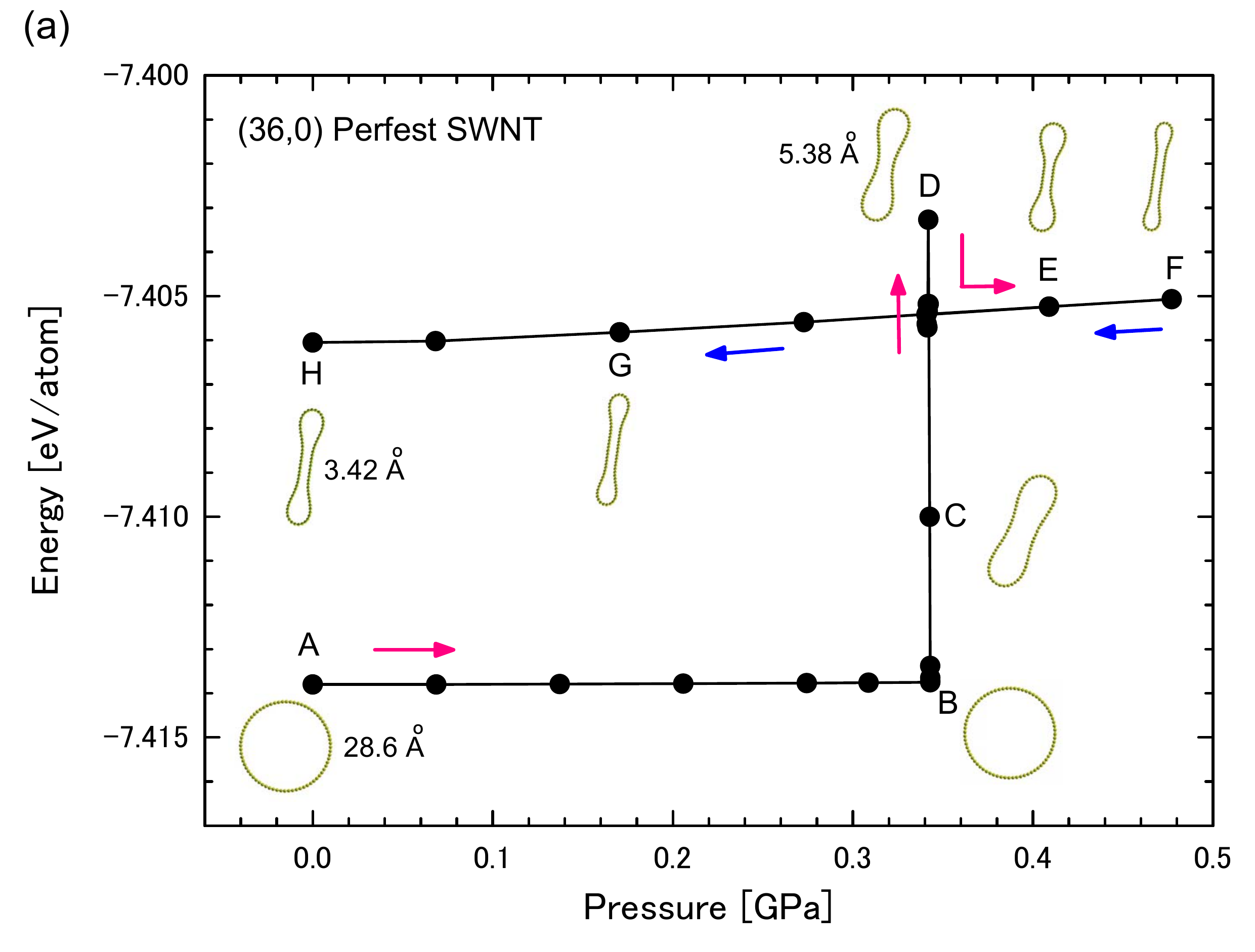}
\includegraphics[width=8.7cm]{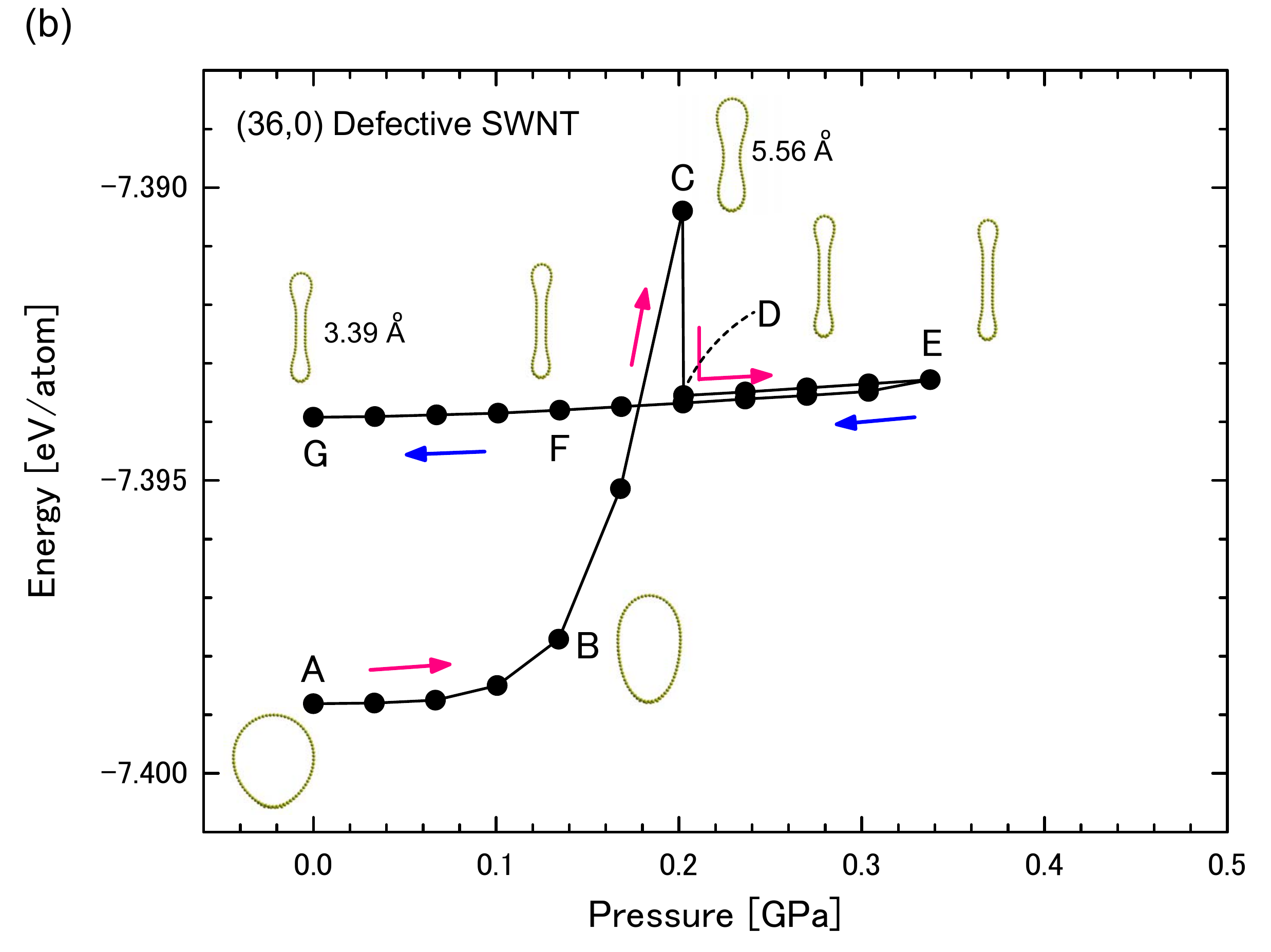}
\caption{Deformation pathway of (36,0) SWNTs under hydrostatic pressure.
(a) Pristine SWNT with no defect. 
The load was increased from the initial state A to the most-squeezed state F 
adiabatically in the alphabetic order,
then decreased from F to the fully-unloaded state H adiabatically, too.
The numeric data in unit of \AA$\;$ indicate the interlayer distance at the closest point in the cross-section.
(b) Defective SWNT with a SW defect. The load was adiabatically imposed in a similar manner to the case of (a).}
\label{fig_swnt}
\end{figure}

\subsection{Hydrostatic pressure application}

In principle, application of external hydrostatic pressure to a CNT
implies that forces are exerted on all the constituent carbon atoms
in the normal direction of the outmost wall.
In practice, however, careful consideration is requisite for 
determining the normal direction of the wall at a given atom,
since the wall is not a continuum layer but a discrete atomic network
and thus there is no unique definition of the direction normal to the wall.
In our simulations, the direction normal to wall is defined by the following procedure \cite{UmenoPE2014}.

Since an atom in the honeycomb-shaped 
carbon network has three nearest neighbors, we find three 
isosceles triangles consisting of the atom and its two neighbors. 
We next define planes on which the three atoms lie and 
its normal vector. Then, a force vector parallel to the 
normal vector is applied to the three atoms that form the isosceles 
triangle. In this way, an atom receives three force vectors combined, 
each of which corresponds to the three planes on which the 
atom sits. 
The vector sum of the three force vectors determines
the direction which we regard as the normal-to-wall direction at the given atom.
For instance, when force vectors with a norm of 0.01 nN
were assigned to the three atoms on the vertices of each isosceles triangle, 
they result in a load of 0.03 nN on an atom in the outmost graphene layer
(note that each atom belongs to vertices of nine triangles).

\subsection{Structural relaxation}

The structural relaxation was performed in the following manner.
First, structural perturbation was given to the initial configuration of 
concentric nanotube cylinders by MD calculation 
at a temperature ($T$) of 100 K for around 100 steps with the time 
step ($\Delta t$) of 2.0 fs. Then, the structure was relaxed employing the 
FIRE algorithm \cite{BitzekPRL2006} until all forces exerted on atoms became less 
than 0.01 eV/\AA. To rule out the possibility of the system falling into
an unstable equilibrium, another MD run at $T = 10$ K was performed 
for around 100 steps with $\Delta t = 2.0$ fs and then the structure was
relaxed again with the same tolerance for forces as above. The
conditions for the initial perturbation were chosen so that the 
nanotube with cylindrical structure was slightly nudged 
to optimize its structure, because otherwise the structure would stay 
unchanged at the unstable equilibrium.

\section{Deformation pathway}

\subsection{Radial deformation of SWNTs}

Figure \ref{fig_swnt} shows the pathway of radial deformation of (36,0) SWNTs 
under hydrostatic pressure.
The vertical axis represents the mechanical energy of the system
calculated by Eq.~(\ref{eq_005}),
and the horizontal axis indicates the pressure applied to the outer cylindrical wall.
The panels of (a) and (b) correspond to a pristine SWNT without defect and 
a defective SWNT containing one SW defect, respectively.
The line graph in the plot is a loading-unloading curve along which
radial deformation proceeds in the alphabetic order;
both the loading (indicated by arrows in magenta) and 
unloading (blue) is adiabatically imposed to the system.
Sequential variation in the cross-section during the loading-unloading process
is also illustrated in the plot.
Numeric data in unit of \AA, attached to a few cross-section views,
show the interlayer distance,
{\it i.e.,} the distance of the two opposed walls at the closest point;
for the circular cross-section, it merely means the diameter of the wall.

It follows from Fig.~\ref{fig_swnt}(a) that a pristine (36,0) SWNT 
with no defect shows an abrupt change in the energy at a critical state,
{\it i.e.,} at the state B in the plot.
Across the state B, the system undergoes the transition 
from the initial circular cross-section to a radially deformed one,
which causes a sudden increase in the energy 
while the applied pressure is kept to be constant.
The energy keeps rising until the system arrives at another critical state,
labeled by D in Fig.~\ref{fig_swnt}(a).
Afterwards, the energy starts to decline suddenly,
even though the magnitude of radial deformation continues to increase.
Further loading results in a most-squeezed state F.
At this extreme state, decompression starts and then the pressure is continued to be reduced adiabatically 
until the fully-unloaded state H.
It should be noted that 
even after the load is completely released,
the initial circular cross-section is not recovered.
This implies that the nanotube collapse is stabilized
due to a certain mechanism, as discussed later.

\begin{figure}[ttt]
\centering
\includegraphics[width=8.7cm]{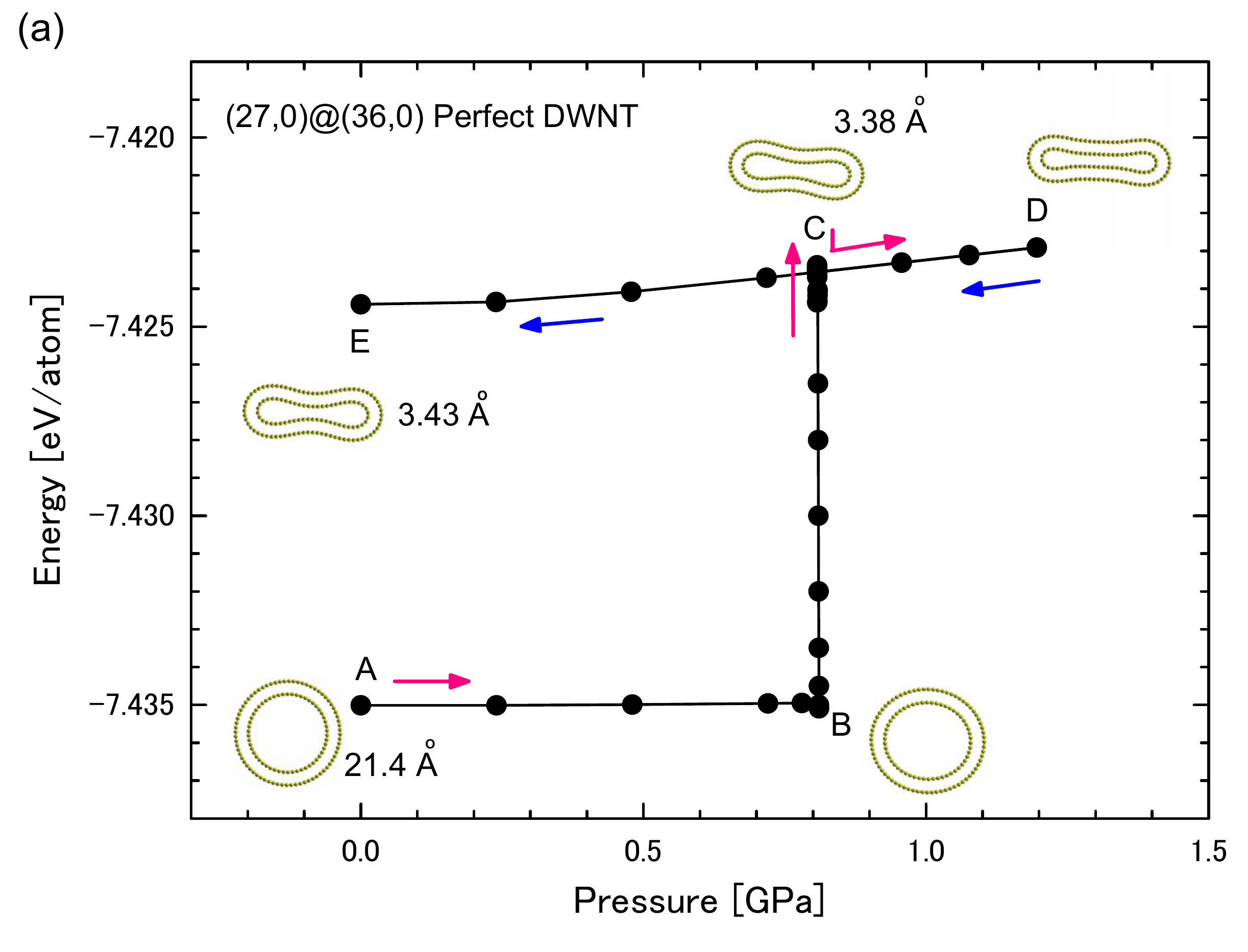}
\includegraphics[width=8.7cm]{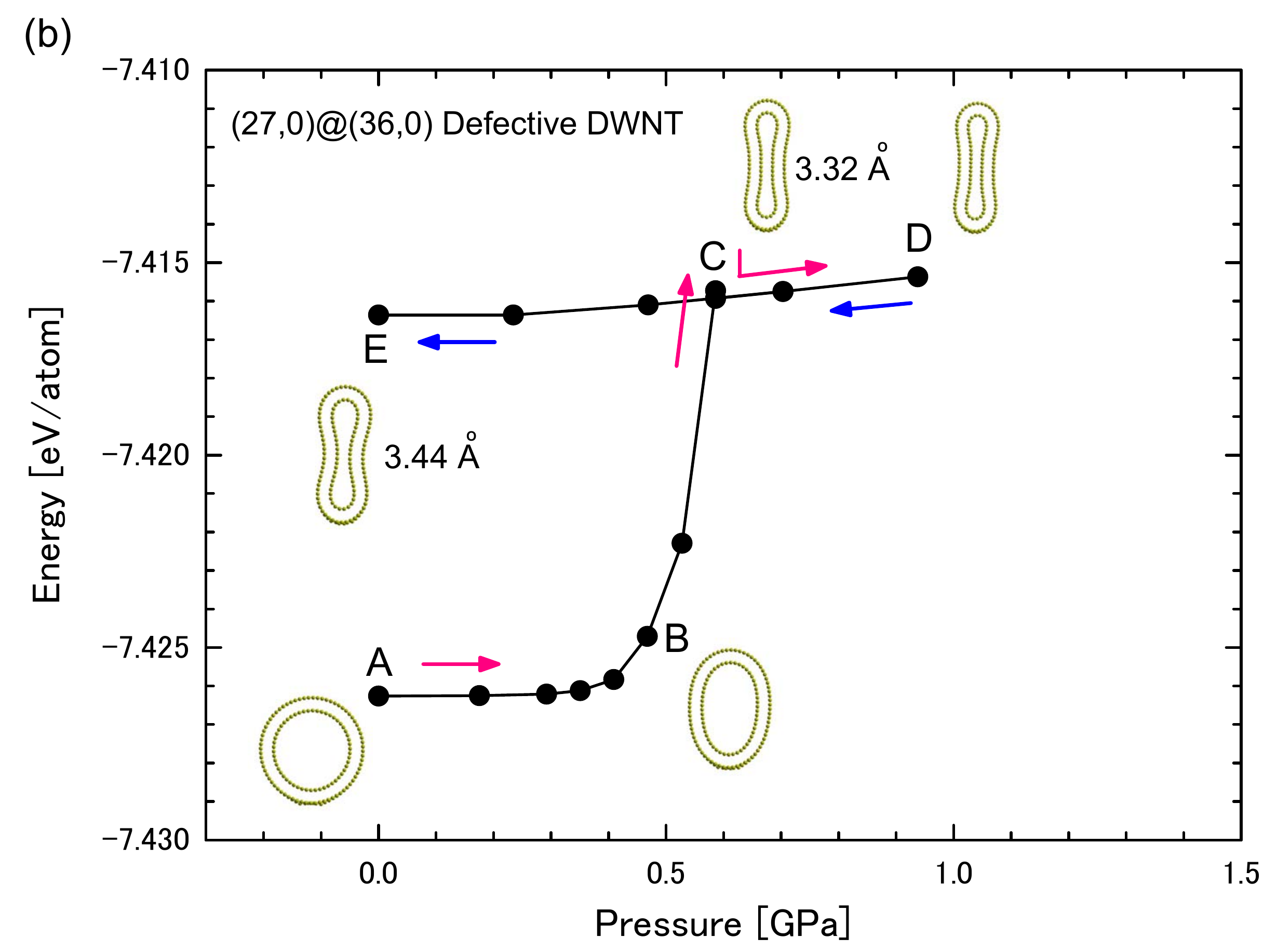}
\caption{Deformation pathway of (27,0)@(36,0) DWNTs under hydrostatic pressure.
(a) Pristine DWNT with no defect. (b) Defective DWNT having one SW defect.
Pressure was varied adiabatically in the alphabetic order.}
\label{fig_dwnt}
\end{figure}

A similar deformation pathway was observed 
for the (36,0) defective SWNT with a SW defect,
as shown in Fig.~\ref{fig_swnt}(b).
The most important difference from the pristine-SWNT case
is the absence of the first critical state
at which the radial buckling occurs.
Instead, the defective SWNT shows a gradual change in the cross-section
from the initial state A to the highest energy state C.
Beyond the state C,
the system undergoes a similar deformation process
to that in Fig.~\ref{fig_swnt}(a).

\subsection{Radial deformation of DWNTs}

Figure \ref{fig_dwnt} shows the deformation pathway
of (27,0)@(36,0) DWNTs;
the panel (a) shows the result of a pristine DWNT,
and the panel (b)  shows the result of a defective DWNT
in which one SW defect is embedded in the outer wall.
As similar to the case of SWNTs,
defect-induced vanishing of the radial buckling is observed in the defective DWNT.
It is also found that the value of pressure requisite for collapsing 
is larger than that in the SWNT case.
This is essentially because a DWNT can be viewed as a SWNT filled by a thinner SWNT.
As such, it is reasonable that the inner tube provides mechanical support against
radial compression \cite{AlencarCarbon2017}
and thus a DWNT should show higher mechanical stability than a SWNT.

A noticeable difference from the case of SWNTs is the suppression of the overshoot 
in the loading-unloading curve near the state C shown in Figs.~\ref{fig_dwnt}(a) and \ref{fig_dwnt}(b).
This suppression can be attributed to lowering of the height in an activation barrier 
that separates elastic and plastic deformation phases,
as will be discussed later.

\section{Energy variation during radial deformation}

\begin{figure}[ttt]
\centering
\includegraphics[width=8.7cm]{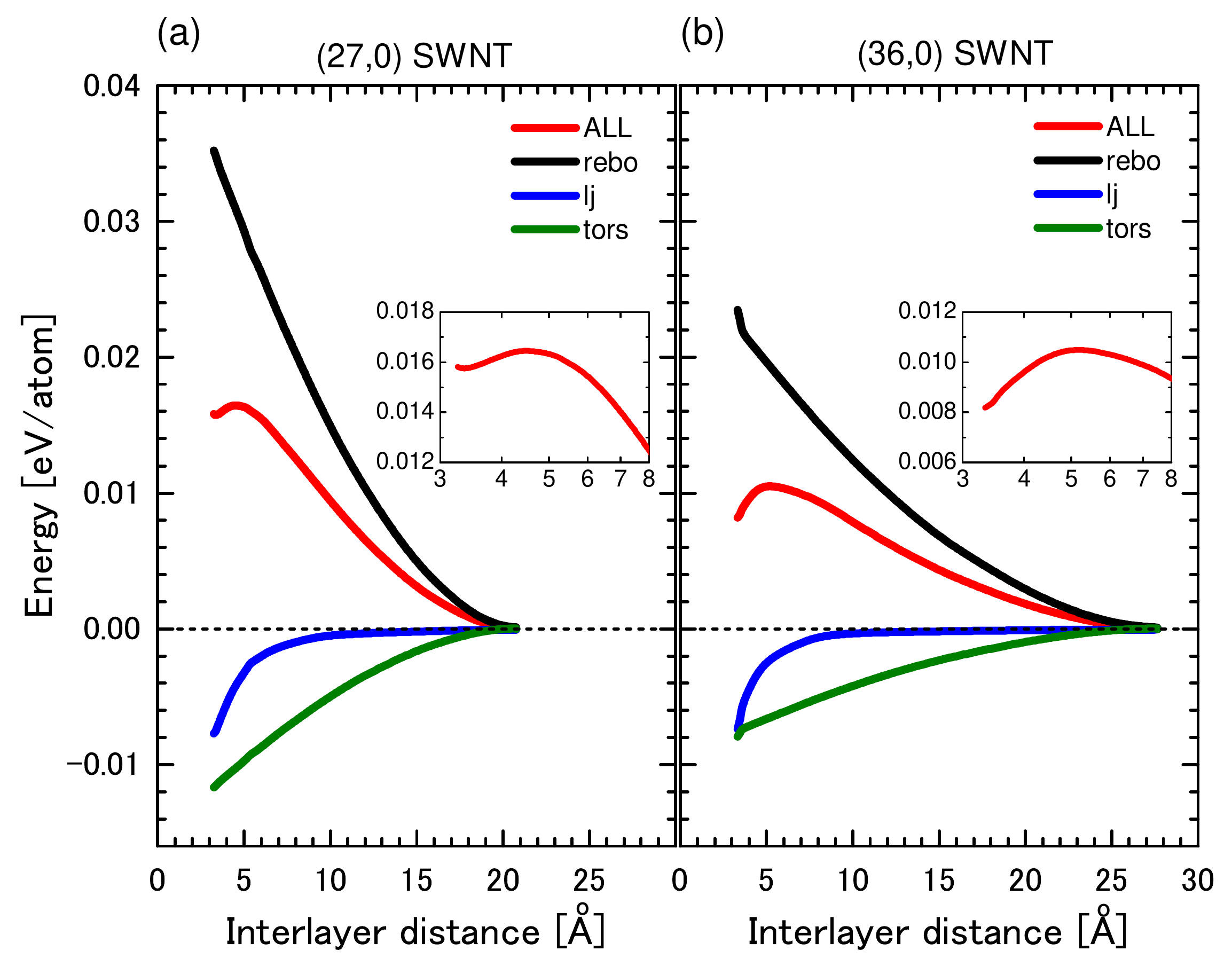}
\includegraphics[width=8.7cm]{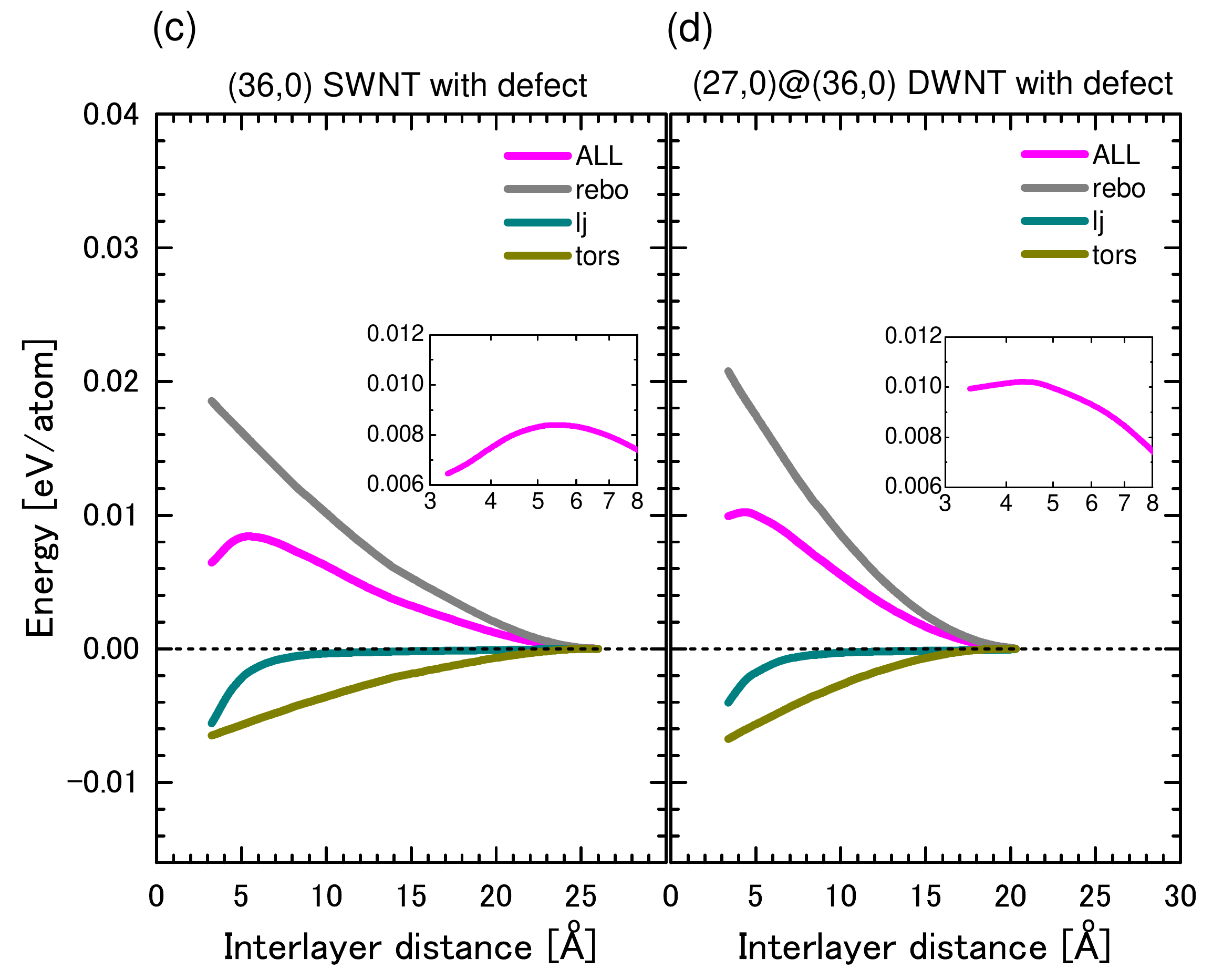}
\caption{AIREBO-derived mechanical energy of pressurized nanotubes.
(a) Pristine (27,0) SWNT; (b) Pristine (36,0) SWNT; (c) Defective (36,0) SWNT;
(d) Defective (27,0)@(36,0) DWNT.
Contributions from three relevant terms contained in AIREBO
are plotted in a separated manner.
Inset: Enlarged view of the total energy curve near the shallow peak.}
\label{fig_component}
\end{figure}

\subsection{Contributions from different AIREBO components}

Figure \ref{fig_component} shows the evolution in the mechanical energy of pressurized SWNTs and DWNTs
with progress of reducing the interlayer distance.
The horizontal axis shows the interlayer distance of the system that tends to decline
as the radial compression proceeds.
The vertical axis is the mechanical energy of the system
calculated on the basis of the AIREBO potential scheme;
contributions from three terms in Eq.~(\ref{eq_005})
are each individually plotted as explained in the legend.
A pristine (27,0) SWNT is also considered together with the nanotube systems
discussed in Figs.~\ref{fig_swnt} and \ref{fig_dwnt},
in order to investigate the effect of tube radius on the energy evolution.

\begin{figure}[ttt]
\centering
\includegraphics[width=8.8cm]{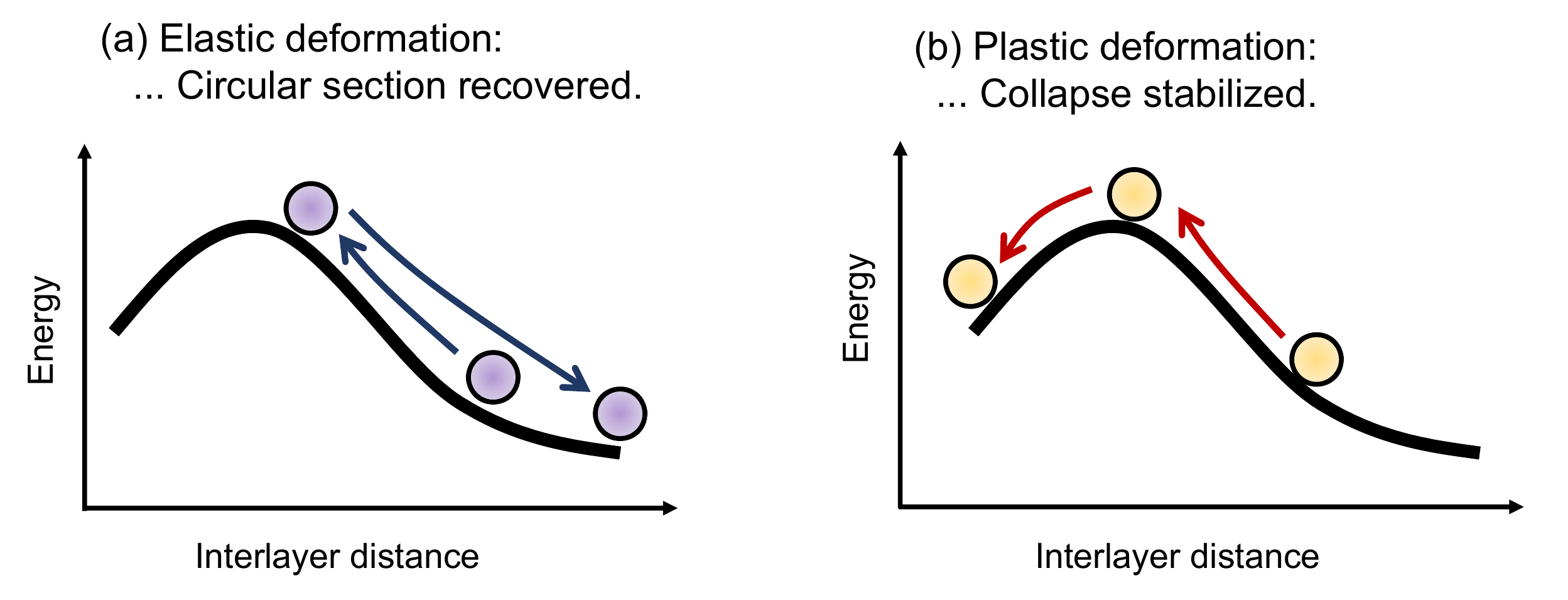}
\caption{Diagram of energy evolution near the activation barrier.
(a) When the magnitude of radial deformation is not sufficient for the system
to overcome the barrier, then the initial circular cross-section will be recovered
after full unloading.
(b) If the cross-section is significantly squeezed and the mechanical energy is raised
enough to overcome the barrier,
then the collapsed cross-section is stabilized even after full unloading.}
\label{fig_barrier}
\end{figure}

In all the plots in Fig.~\ref{fig_component},
it is commonly observed that
the total energy starts to increase in a monotonic manner 
with decreasing interlayer distance,
followed by a slight drop.
In the vicinity of ca. 5 \AA,
the total energy curve shows a convex shape,
as illustrated by the magnified view in the insets.
This convex energy profile plays a role of the activation barrier
beyond which the collapsed cross-section is stabilized;
see Fig.~\ref{fig_barrier} for schematic illustrations
of energy evolution near the activation barrier.
Given an insufficient radial deformation,
the system can not reach a full collapse state but return to
the initial circular cross-section after full unloading.
In contrast, if the cross-section is significantly squeezed and the mechanical energy is raised
enough to overcome the activation barrier,
then the collapsed cross-section is stabilized even after full unloading

It is interesting to note that the interlayer distance which corresponds to the maximum position
[{\it e.g.,} 5.3 \AA$\;$ for pristine (36,0) SWNT ]
is nearly equal to that realized in the second critical state
in the deformation pathway
[5.38 \AA$\;$ for pristine (36,0) SWNT as marked at the state D in Fig.~\ref{fig_swnt}(a)].
This implies that the overshoot height in the loading-unloading curve
is determined by the activation barrier height
estimated from the energy evolution curve in Fig.~\ref{fig_component}.

\subsection{Mechanism of negative contribution from the torsion term}

The most important observation in Fig.~\ref{fig_component} is
the monotonically decreasing trend in the energy curve derived from the torsion term.
As followed from Fig.~\ref{fig_component},
the torsion term as well as the vdW term
give negative contributions at all the interlayer distance,
and their magnitudes are enhanced as the collapse advances.
In contrast, the REBO term gives positive contribution at every interlayer distance.
The sum of the positive and negative contributions 
results in the convex shape in the total energy curve.

We emphasize here that
the absolute value of the torsion term contribution (negative)
is always greater than that of the vdW term contribution (negative),
as least in the range of interlayer distance we have considered.
This fact indicates that the interatomic interaction described by the torsion term 
is fundamental for stabilizing the collapsed cross-section,
together with the attractive interaction driven by the vdW forces
between the two opposing walls.
To the knowledge of the authors, 
the contribution from the torsion term to the radial deformation of carbon nanotubes
has never been examined in detail thus far, 
compared with the intense studies on the counterparts
regarding the REBO term and vdW term.
In this sense, this is the first numerical evidence
that the torsion term plays a significant role in the stabilization
of collapsed cross-section of carbon nanotubes.

\begin{figure}[ttt]
\centering
\includegraphics[width=8.2cm]{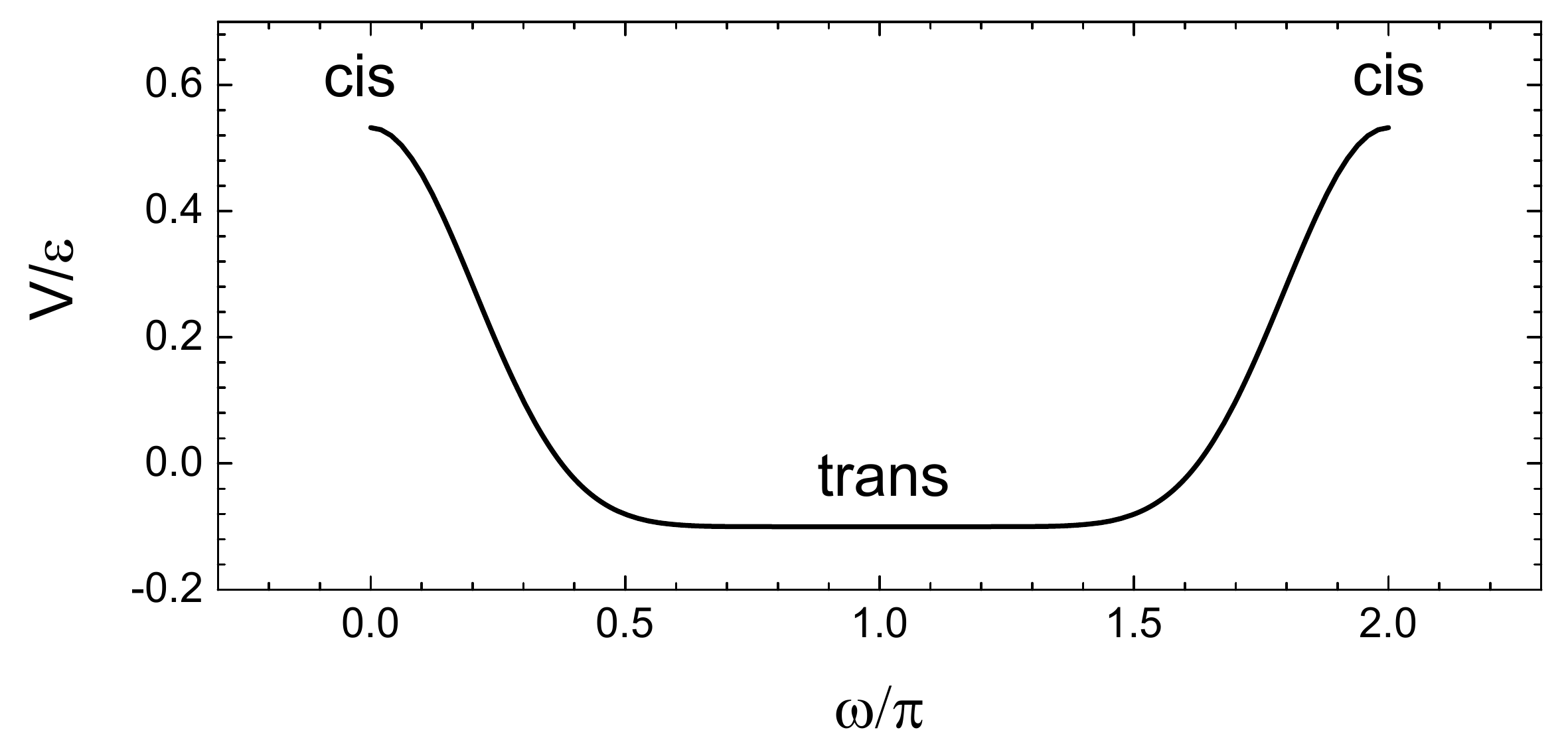}
\caption{Profile of the normalized potential $V^{\rm tors}/\varepsilon_{kji\ell}$ 
with regard to the dihedral angle $\omega_{kji\ell}$ (abbreviated by $\omega$).
The upward sharp peak at $\omega=0$ (or $2\pi$) indicates that
the {\it cic}-geometry is energetically unstable for a four-atomic carbon chain.
In contrast, the downward plateau around $\omega=\pi$ indicates
that the {\it trans}-geometry is energetically favored and 
rather robust against twist deformation of the chain.}
\label{V_potential_profile}
\end{figure}

The reason why the torsion term decreases monotonically with approaching the interlayer distance
is explained by considering the dependence of $V^{\rm tors}$,
defined by Eq.~(\ref{eq_125}),
on the dihedral angle $\omega_{kji\ell}$.
Figure \ref{V_potential_profile}
shows the profile of $V^{\rm tors}$ as a function of $\omega_{kji\ell}$.
We see from the figure that 
$V^{\rm tors}$ has 
a relatively sharp upward peak at $\omega_{kji\ell}=0$,
which corresponds to {\it cis}-type geometry of the $k$-$j$-$i$-$\ell$ carbon atomic chain
that was depicted in Fig.~\ref{fig_bonding}(b).
In addition, $V^{\rm tors}$ has
a very broad minimum around $\omega_{kji\ell}=\pi$,
which corresponds to {\it trans}-type geometry of the $k$-$j$-$i$-$\ell$ chain.

An important feature of the profile is the significant asymmetry in the peak width 
between the upward peak at $\omega=0$ (sharp) and the plateau-like downward peak
around $\omega=\pi$ (quite broad).
The upward peak at $\omega=0$ indicates that {\it cis}-type geometry of a four-atom chain
is energetically unstable and thus a feasible decline in the mechanical energy
of chain will be observed if it is twisted ({\it i.e.,} if $\omega$ deviates from 0).
In contrast, the downward plateau around $\omega=\pi$ indicates that
a {\it trans}-type chain will show only a slight change in energy
even when it is twisted.
Therefore, if both {\it cis}- and {\it trans}-type chains are twisted simultaneously
with the same twist angle, then the energy decrease caused by {\it cis} deformation will overcome
the energy increase by {\it trans} deformation,
and thus the total energy of the two chains will be reduced.

The above argument regarding a pair of {\it cis}- and {\it trans}-type chains
holds true for a monoatomic carbon cylinder ({\it i.e.,} a SWNT),
in which many {\it cis}- and {\it trans}-type chains are embedded.
Under cross-sectional deformation of a SWNT, therefore,
accumulation of local energy reduction derived from many twisted {\it cis}-chains
will overcome a slight energy increase derived from many twisted {\it trans}-chains.
As a result, $E_{kij\ell}^{\rm tors}$ of the system will decrease as the cross-sectional deformation proceeds.

Our finding provides a new insight into the atomistic-scale energetics of collapsing carbon nanotubes.
It has been conventionally thought that the vdW force is a primary (and perhaps the only) driving force
that stabilizes fully collapsed cross-sections, because the attractive nature of vdW interaction between to opposing carbon walls
may result in high mechanical stability of the full collapse.
But as clearly demonstrated in Fig.~\ref{fig_component},
the monotonic decline in the torsion-induced energy curve
should be also indispensable for realizing the stabilization of the full collapse,
taking a complementary role with the attractive vdW interaction.
This is the main finding of this article.

\subsection{Crossover between elastic and plastic deformation phases}

We have seen that there is an activation barrier for stabilizing collapsed cross-sections.
If the pressure-induced reduction in the interlayer distance at the closest point is not sufficiently
small, then the system behaves elastically so that the initial circular cross-section will be recovered
after complete unloading [see Fig.~\ref{fig_barrier}(a)].
In contrast, once the cross-section is squeezed sufficiently, 
then the system undergoes plastic deformation;
in this case, the collapsed cross-section is stabilized even after complete unloading
[see Fig.~\ref{fig_barrier}(b)].
In this context, it is interesting to probe the threshold interlayer distance
that separates the elastic and plastic deformation phases.

\begin{figure}[ttt]
\centering
\includegraphics[width=8.7cm]{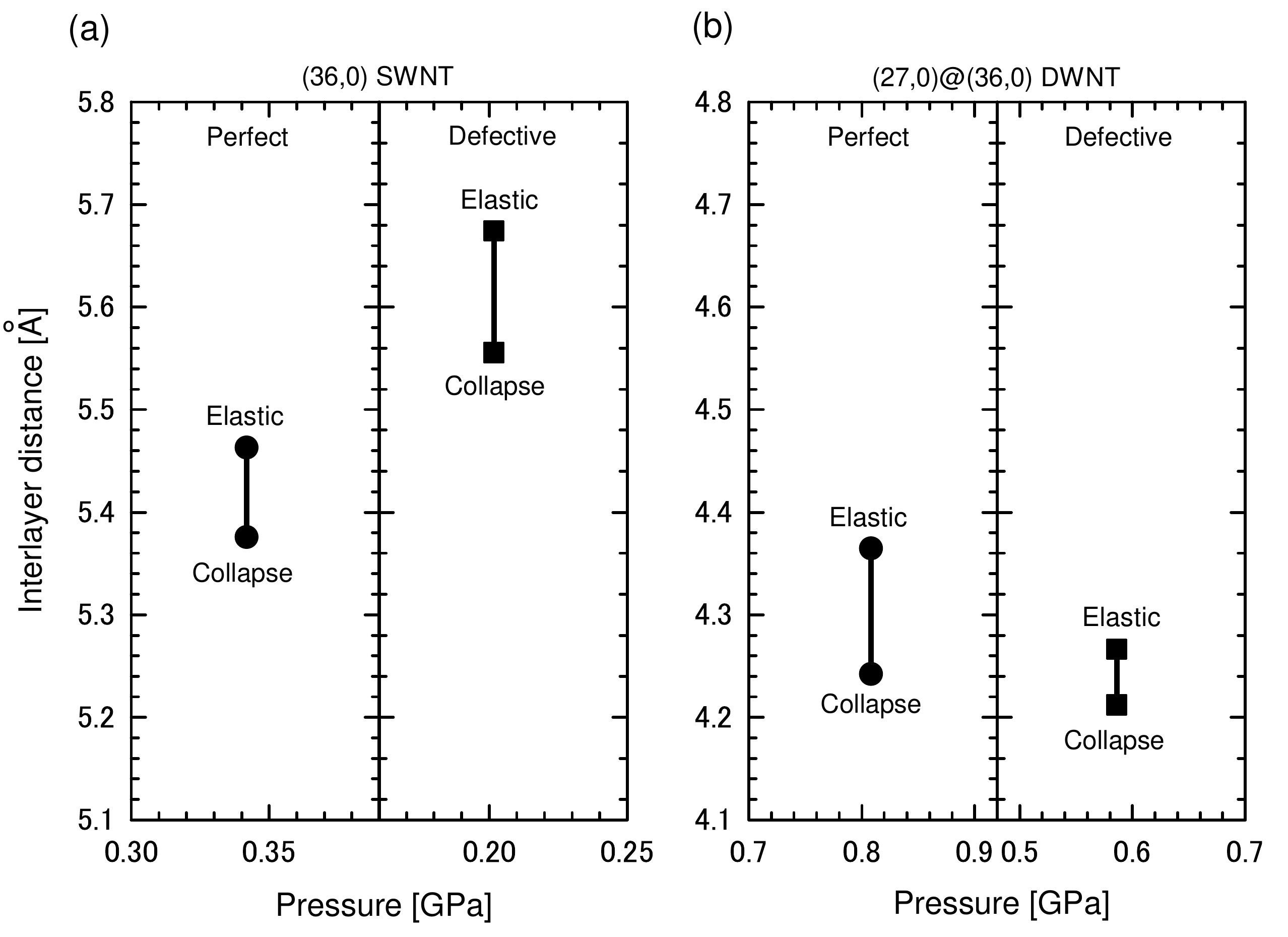}
\caption{Diagram of the threshold interlayer distance (ID)
that separates elastic and plastic deformation phases.
Each vertical segment indicates the marginal region of ID.
If radial compression continues until ID is reduced below the bottom edge of the segment,
then the system will fall into a full collapse 
even after complete unloading [see Fig.~\ref{fig_barrier}(a)].
If radial compression stops before ID goes below the upper edge of the segment,
then the system will recover the initial circular cross-section 
after complete unloading [see Fig.~\ref{fig_barrier}(b)].
}
\label{fig_threshold}
\end{figure}

Figure \ref{fig_threshold} presents a diagram of the threshold interlayer distance (ID)
that separates elastic and plastic deformation phases.
Each vertical segment indicates the marginal region of ID.
If radial compression continues until ID is reduced below the bottom edge of the segment,
then the system will fall into a full collapse 
even after complete unloading [see Fig.~\ref{fig_barrier}(a)].
If radial compression stops before ID goes below the upper edge of the segment,
then the system will recover the initial circular cross-section 
after complete unloading [see Fig.~\ref{fig_barrier}(b)].

Figure \ref{fig_threshold}(a) indicates that for SWNT,
SW defect insertion leads to upward shifts both in the two threshold interlayer distances,
{\it i.e.,} the bottom limit of the elastic deformation phase
and the upper limit of the plastic deformation phase.
These upward shifts are intuitively understood
as the defect insertion may cause reduced softening and thus the collapse becomes easy to occur
even at a relatively large interlayer distance compared with the defect-free counterpart.
But this softening effect is not significant for DWNTs as shown in Fig.~\ref{fig_threshold}(b).
This is because in the defective DWNT,
it is the outer wall that contains a defect;
the inner wall is free from the defect
and thus tends to push back the outer wall under radial compression.

\section{Conclusion}

We have conducted AIREBO-based MD simulations of collapsing SWNTs and DWNTs
under hydrostatic pressure.
The main finding is that the torsion term, one of the constituent potential terms in AIREBO,
takes a vital role in stabilizing fully collapsed cross-section of carbon nanotubes.
The novelty of this result stems from the fact that
in established understandings, the vdW term is quite often considered 
as the only potential term that stabilizes collapsed cross-sections.
Our simulation data clearly demonstrate that 
the sum of decreases both in the torsion term and the vdW term
with progress of radial compression,
not only the contribution from the vdW term,
is the physical origin of carbon nanotube collapse.
It was also found that the above conclusion holds true
no matter whether the system contains a SW defect.

\section*{Acknowledgments}
This work was supported by JSPS KAKENHI Grant Numbers 
JP 15H03888, 15H04207, and 15KK0220.

\bibliographystyle{elsarticle-num}
\bibliography{Umeno_Torsion_201806.bib}

\end{document}